# Level Statistics of $SU(3) \leftrightarrow \overline{SU(3)}$ Transitional Region

M. A. Jafarizadeh[a,b] [*], N. Fouladi[c], H. Sabri[c†], P. Hossein nezhade gavifekr [c], Z. Ranjbar [c]

[a]Department of Theoretical Physics and Astrophysics, University of Tabriz, Tabriz 51664, Iran.

[b]Research Institute for Fundamental Sciences, Tabriz 51664, Iran.

[c] Department of Nuclear Physics, University of Tabriz, Tabriz 51664, Iran.

---

[*] E-mail: jafarizadeh@tabrizu.ac.ir
[†] E-mail: h-sabri@tabrizu.ac.ir

## Abstract

The level statistics of $SU(3) \leftrightarrow \overline{SU(3)}$ transitional region of IBM is described by the nearest neighbor spacing distribution statistics. The energy levels are determined by using the $SO(6)$ representation of eigenstates. By employing the MLE technique, the parameter of Abul-Magd distribution is estimated where suggest less regular dynamics for transitional region in compare to dynamical symmetry limits. Also, the $O(6)$ dynamical symmetry which is known as the critical point of this transitional region, describes a deviation to more regular dynamics.

## Introduction

The investigation of significant changes in energy levels and electromagnetic transition rates resulting in the shape phase transitions [1-4] from one kind of collective behavior to another has received a lot of attention in recent years. The new symmetries called $X(5)$ and $E(5)$ are obtained within the framework of the collective model have employed to describe atomic nuclei at the critical points [5-7]. In the interacting boson model (IBM) framework [8-11], a very simple two-parameter description has been used leading to a symmetry triangle describing many atomic nuclei. This model describes the nuclear structure of the even–even nuclei within the $U(6)$ symmetry, possessing the $U(5)$, $SU(3)$ and $O(6)$ dynamical symmetry limits. No phase transition is found between the $SU(3)$ and $O(6)$ vertices of the triangle. However, as discussed in Refs. [12-13] in the context of catastrophe theory, an analysis of the separatrix of the IBM-1 Hamiltonian in the coherent state formalism shows that there is a phase transition in between oblate ($\overline{SU(3)}$) and prolate ($SU(3)$) deformed nuclei. This phase transition and its critical point symmetry, which in fact, coincides by the O(6) limit has described from the standpoint of physical observables in Refs.[14-16] by Jolie *et al*.

On the other hand, the energy level statistics [17-19] has employed as a new observable to analyze the evolution of energy spectra along the two dynamical symmetry limits. It's well-known [20-21], the energy spectra of nuclei correspond to any definite phase, governed due to the dynamical symmetry limits, exhibit more regularity. On the other hand, in the transitional region which contains the critical point of quantum phase transition, system goes from a symmetry limit to another dynamical symmetry limit. Consequently, a combination of different symmetries visualizes by nuclei and a deviation of regular dynamics occurs in spectra.

In the present paper, we investigate the fluctuation properties of energy spectra for $SU(3) \leftrightarrow \overline{SU(3)}$ transitional region in the Nearest Neighbor Spacing Distribution (NNSD) statistics framework. To determine the energy spectra of considered region, the $SO(6)$ representation of eigenstates [22-23] are generalized for systems with total boson number $N = 3, 4$ and the energy eigenvalues determined by parameter-free techniques. With using these energy spectra, sequences are constructed by unfolding procedure and then, by employing the Maximum Likelihood Estimation (MLE) technique, the parameter of Abul-Magd's distribution estimated [24-25]. The ML-based estimated values propose an apparent dependence between the chaocity of considered sequences and control parameter where suggest a deviation to less regular dynamics for transitional region in compare to both prolate and oblate dynamical symmetry limits. Also, the $\chi = 0$ which is correspond to $O(6)$ dynamical symmetry limits, namely the critical point of this transitional region, suggest more regular dynamics in compare to other values of control parameter.

This paper is organized as follows: section 2 briefly summarizes the theoretical aspects of considered Hamiltonian and $SO(6)$ representations of eigenstates. In section 3, details about statistical investigation are presented which includes unfolding procedure and MLE technique which applied to Abul-Magd distribution. Numerical results are presented in Section 4. Section 5 is devoted to summarize and some conclusion based on the results given in section 4.

## 2. Theoretical framework

The phase transition have been studied widely in Refs.[14-16], are those of the ground state deformation. In the Interacting Boson Model (IBM), one would achieve a very simple two parameters description leading to a symmetry triangle which is known as extended Casten triangle [11]. There are four dynamical symmetries of the IBM called $U(5)$, $SU(3)$, $\overline{SU(3)}$ and $O(6)$ limits. They correspond to vibrational nuclei with a spherical form, namely $U(5)$, an axially symmetric prolate rotor with a minimum in the energy at $\gamma = 0°$ which corresponds to $SU(3)$ and an axially symmetric oblate rotor with a minimum at $\gamma = 60°$, namely $\overline{SU(3)}$. The fourth symmetry is located in the middle of the $SU(3) \leftrightarrow \overline{SU(3)}$ transitional region and corresponds to a rotor with a flat potential in $\gamma$, it means $O(6)$ limit, as have presented in Figure1. It is parameterized using the simple Hamiltonian [14-15]

$$\hat{H}(N,\eta,\chi) = \eta \hat{n}_d + \frac{\eta-1}{N}\hat{Q}_\chi \hat{Q}_\chi \quad , \tag{2.1}$$

Where $\hat{n}_d = d^\dagger \cdot \tilde{d}$ is the $d$–boson number operator and $\hat{Q}_\chi = (s^\dagger \tilde{d} + d^\dagger s)^{(2)} + \chi(d^\dagger \times \tilde{d})^{(2)}$ represents the quadrupole operator and $N(=n_s + n_d)$ stands for the total number of bosons. The $\eta$ and $\chi$ regard as control parameters while vary within the range $\eta \in [0,1]$ and $\chi \in [-\sqrt{7/2}, +\sqrt{7/2}]$. Our considered region, namely the prolate-oblate transitional region, passing through the $O(6)$ dynamical symmetry limit, is known to be situated close to the upper right leg of the extended Casten triangle with $\eta = 0$. In the following, we have employed the $SO(6)$ representation to determine the eigenvalues of Hamiltonian (2.1). The Algebraic structure of IBM has been described in detail in Refs.[22-23]. Here, we briefly outline the basic ansatz and summarize the results have obtained in this paper for our considered representation. The classification of states in the IBM $SO(6)$ limit is [26-27]

$$U(6) \supset SO(6) \supset SO(5) \supset SO(3) \supset SO(2) \quad , \tag{2.2}$$
$$\downarrow \quad \downarrow \quad \downarrow \quad \downarrow \quad \downarrow$$
$$[N] \quad \langle \Sigma \rangle \quad (\tau) \quad L \quad M$$

The multiplicity label $\nu_\Delta$ in the $SO(5) \supset SO(3)$ reduction will be omitted in the following when it is not needed. The eigenstates $|[N]\langle\sigma\rangle(\tau)\nu_\Delta LM\rangle$ are obtained with a Hamiltonian with the $SO(6)$ dynamical symmetry. The construction of our considered representation requires n-boson creation and annihilation operators with definite tensor character in the basis (2.2) as;

$$B^\dagger_{[N]\langle\sigma\rangle(\tau)lm} \quad , \quad \tilde{B}_{[n^5]\langle\sigma\rangle(\tau)lm} \equiv (-1)^{l-m}(B^\dagger_{[N]\langle\sigma\rangle(\tau)l,-m})^\dagger \quad , \tag{2.3}$$

Of particular interest are tensor operators with $\sigma < n$. They have the property

$$\tilde{B}_{[n^5]\langle\sigma\rangle(\tau)lm}|[N]\langle N\rangle(\tau)\nu_\Delta LM\rangle = 0 \quad , \quad \sigma < n \tag{2.4}$$

For all possible values of $\tau$ and $L$ contained in the $SO(6)$ irrep $\langle N \rangle$. This is so because the action of $\tilde{B}_{[n^5]\langle\sigma\rangle(\tau)lm}$ leads to an $(N-n)$–boson state which contains the $SO(6)$ irrep $\langle \Sigma \rangle = \langle N-n-2i \rangle$, $i = 0,1,...$, which cannot be coupled with $\langle\sigma\rangle$ to yield $\langle\Sigma\rangle = \langle N \rangle$, since $\sigma < n$. Number conserving normal ordered interactions that are constructed out of such tensors with $\sigma < n$ (and their Hermitian conjugates) thus have $|[N]\langle N\rangle(\tau)\nu_\Delta LM\rangle$ as eigenstates with zero eigenvalues. A systematic enumeration of all interactions with this property is a simple matter of $SO(6)$ coupling. For one body operators,

$$B^\dagger_{[1]\langle 1\rangle(0)00} = s^\dagger \equiv b^\dagger_0 \quad , \quad B^\dagger_{[1]\langle 1\rangle(1)2m} = d^\dagger_m \equiv b^\dagger_{2m} \quad , \tag{2.5}$$

On the other hand, coupled two body operators are of the form

$$B^\dagger_{[2]\langle\sigma\rangle(\tau)lm} \propto \sum_{\tau_k \tau_{k'}} \sum_{kk'} C^{\langle\sigma\rangle(\tau)l}_{\langle 1\rangle(\tau_k)k,\langle 1\rangle(\tau_{k'})k'} (b^\dagger_k b^\dagger_{k'})^{(l)}_m \quad , \tag{2.6}$$

Where $(b^\dagger_k b^\dagger_{k'})^{(l)}_m$ represent coupling to angular momentum $(l)$ and the $C$ coefficients are known $SO(6) \supset SO(5) \supset SO(3)$ isoscalar factors [28]. These processes lead to the normalized two-boson $SO(6)$ representation displayed in Tables (1-2) for systems with total boson number $N = 3$ and 4, respectively.

Table1. The $SO(6)$ representation of eigenstates for systems with total boson number $N (= n_s + n_d) = 3$.

| $n_d$ | $\sigma$ | $\tau$ | $l$ | Representation |
|---|---|---|---|---|
| 3 | 3 | 3 | 6 | $\sqrt{1/6}[(d^\dagger \times d^\dagger)^4 \times d^\dagger]^6_m$ |
| 3 | 3 | 3 | 4 | $\sqrt{7/22}[(d^\dagger \times d^\dagger)^2 \times d^\dagger]^4_m$ |
| 3 | 3 | 3 | 3 | $\sqrt{7/30}[(d^\dagger \times d^\dagger)^2 \times d^\dagger]^3_m$ |
| 3 | 1 | 1 | 2 | $\sqrt{5/14}[(d^\dagger \times d^\dagger)^0 \times d^\dagger]^2_m$ |
| 3 | 3 | 3 | 0 | $\sqrt{1/6}[(d^\dagger \times d^\dagger)^2 \times d^\dagger]^0_0$ |
| 2 | 3 | 2 | 4 | $\sqrt{1/2}[(d^\dagger \times d^\dagger)^4 \times s^\dagger]^4_m$ |
| 2 | 3 | 2 | 2 | $\sqrt{1/2}[(d^\dagger \times d^\dagger)^2 \times s^\dagger]^2_m$ |
| 2 | 1 | 0 | 0 | $\sqrt{1/2}[(d^\dagger \times d^\dagger)^0 \times s^\dagger]^0_0$ |
| 1 | 3 | 1 | 2 | $\sqrt{1/2}[(d^\dagger \times s^\dagger)^2 \times s^\dagger]^2_m$ |
| 0 | 3 | 0 | 0 | $\sqrt{1/6}[(s^\dagger \times s^\dagger)^0 \times s^\dagger]^0_0$ |

Table2. The $SO(6)$ representation of eigenstates for systems with total boson number $N (= n_s + n_d) = 4$.

| $n_d$ | $\sigma$ | $\tau$ | $l$ | Representation | $n_d$ | $\sigma$ | $\tau$ | $l$ | Representation |
|---|---|---|---|---|---|---|---|---|---|
| 4 | 4 | 4 | 8 | $\sqrt{1/24}[(d^\dagger \times d^\dagger)^4 \times (d^\dagger \times d^\dagger)^4]^8_m$ | 4 | 4 | 4 | 6 | $\sqrt{7/60}[(d^\dagger \times d^\dagger)^4 \times (d^\dagger \times d^\dagger)^2]^6_m$ |
| 4 | 4 | 4 | 5 | $\sqrt{1/12}[(d^\dagger \times d^\dagger)^4 \times (d^\dagger \times d^\dagger)^2]^5_m$ | 4 | 4 | 4 | 4 | $\sqrt{49/664}[(d^\dagger \times d^\dagger)^2 \times (d^\dagger \times d^\dagger)^2]^4_m$ |
| 4 | 2 | 2 | 4 | $\sqrt{5/36}[(d^\dagger \times d^\dagger)^0 \times (d^\dagger \times d^\dagger)^4]^4_m$ | 4 | 2 | 2 | 2 | $\sqrt{5/36}[(d^\dagger \times d^\dagger)^0 \times (d^\dagger \times d^\dagger)^2]^2_m$ |
| 4 | 0 | 0 | 0 | $\sqrt{5/56}[(d^\dagger \times d^\dagger)^0 \times (d^\dagger \times d^\dagger)^0]^0_0$ | 4 | 4 | 4 | 2 | $\sqrt{5/48}[((d^\dagger \times d^\dagger)^2 \times d^\dagger)^0_0 \times d^\dagger]^2_m$ |
| 3 | 4 | 3 | 6 | $\sqrt{1/6}[(d^\dagger \times d^\dagger)^4 \times (d^\dagger \times s^\dagger)^2]^6_m$ | 3 | 4 | 3 | 4 | $\sqrt{7/22}[(d^\dagger \times d^\dagger)^2 \times (d^\dagger \times s^\dagger)^2]^4_m$ |
| 3 | 4 | 3 | 3 | $\sqrt{7/30}[(d^\dagger \times d^\dagger)^2 \times (d^\dagger \times s^\dagger)^2]^3_m$ | 3 | 2 | 1 | 2 | $\sqrt{5/14}[(d^\dagger \times d^\dagger)^0 \times (d^\dagger \times s^\dagger)^2]^2_m$ |
| 3 | 4 | 3 | 0 | $\sqrt{1/6}[((d^\dagger \times d^\dagger)^2 \times d^\dagger)^0_0 \times s^\dagger]^0_0$ | 2 | 4 | 2 | 4 | $\sqrt{1/4}[(d^\dagger \times d^\dagger)^4 \times (s^\dagger \times s^\dagger)^0]^4_m$ |
| 2 | 4 | 2 | 2 | $\sqrt{1/4}[(d^\dagger \times d^\dagger)^2 \times (s^\dagger \times s^\dagger)^0]^2_m$ | 2 | 2 | 0 | 0 | $\sqrt{1/4}[(d^\dagger \times d^\dagger)^0 \times (s^\dagger \times s^\dagger)^2]^0_0$ |
| 1 | 4 | 1 | 2 | $\sqrt{1/6}[(d^\dagger \times s^\dagger)^2 \times (s^\dagger \times s^\dagger)^0]^2_m$ | 0 | 4 | 0 | 0 | $\sqrt{1/24}[(s^\dagger \times s^\dagger)^0 \times (s^\dagger \times s^\dagger)^0]^0_0$ |

Now, with using these eigenstates, one can determine energy spectra for considered systems as

$$\langle [N]\langle\sigma\rangle(\tau)v_\Delta LM | H | [N]\langle\sigma\rangle(\tau)v_\Delta LM \rangle = \eta n_d + \frac{\eta - 1}{N} \varepsilon \quad , \tag{2.7}$$

Where $\varepsilon$ denote the matrix elements of quadrupole term in Hamiltonian as presented in Tables (3-4) for systems with $N = 3, 4$, respectively.

Table3. The elements of quadrupole operator in Hamiltonian (2.1) for systems with $N=3$ which determined by states introduced in Table2.

| $L$ | $\varepsilon$ | $L$ | $\varepsilon$ |
|---|---|---|---|
| 0 | $315 + 360\chi^2 + \frac{720}{7}\chi^4$ | 2 | $187 + \frac{7113}{7}\chi^2 + \frac{5373}{49}\chi^4 + \frac{473}{343}\chi^6$ |
| 3 | $3 + \frac{18}{7}\chi^2$ | 4 | $33 - 8\chi^2 + \frac{396}{49}\chi^4$ |
| 6 | $3 + \frac{33}{7}\chi^2$ | | |

Table4. The elements of quadrupole operator in Hamiltonian (2.1) for systems with $N=4$ which determined by states introduced in Table3.

| $L$ | $\varepsilon$ |
|---|---|
| 0 | $31360 + \frac{480454}{7}\chi^2 + \frac{272538}{7}\chi^4 + \frac{69120}{49}\chi^6$ |
| 2 | $402 + \frac{2056}{7}\chi^2 + \frac{3108}{7}\chi^4 + \frac{3971}{49}\chi^6 + \frac{316}{343}\chi^8 + \frac{471}{2401}\chi^{10}$ |
| 3 | $14 + \frac{18}{7}\chi^2$ |
| 4 | $275 + \frac{1902}{7}\chi^2 + \frac{167}{7}\chi^4 + \frac{1059}{49}\chi^6 + \frac{605}{343}\chi^8$ |
| 5 | $4 + \frac{24}{7}\chi^2$ |
| 6 | $56 + \frac{290}{7}\chi^2 + \frac{1221}{49}\chi^4$ |
| 8 | $4 + \frac{52}{7}\chi^2$ |

Now, we are proceeding to determine the energy spectra by parameter free method, i.e. up to over all scale factors while we have considered all levels in our sequences.

## 3. The method of Statistical analysis

The spectral fluctuations of low-lying nuclear levels have been considered by different statistics such as Nearest Neighbor Spacing Distribution (NNSD) [17], linear coefficients between adjacent spacing [18] and Dyson-Mehta $\Delta_3(L)$ statistics [29-30] which based on the comparison of statistical properties of nuclear spectra with the predictions of Random Matrix Theory (RMT). The NNSD, or $P(s)$ functions, is the observable most commonly used to analyze the short-range fluctuation properties in the nuclear spectra. The NNSD statistics would perform by complete (few or no missing levels) and pure (few or no unknown spin-parities) level scheme [18] where these conditions are available for a limited number of nuclei. Therefore, to obtain the statistically relevant samples, we in need to combine different level schemes to construct sequences. To compare the different sequences to each other, each set of energy levels must be converted to a set of normalized spacing, namely, each sequence must be unfolded. To unfold our spectrum, we had to use some levels with same symmetry. For a given spectrum $\{E_i\}$, it is necessary to separate it into the fluctuation part and the smoothed average part, whose behavior is nonuniversal and cannot be described by RMT [17]. To do so, we count the number of the levels below $E$ and write it as

$$N(E) = N_{av}(E) + N_{fluct}(E) \qquad , \qquad (3.1)$$

Then we fix the $N_a(E_i)$ semiclasically by taking a smooth polynomial function of degree 6 to fit the staircase function $N(E)$. We obtain finally, the unfolded spectrum with the mapping

$$\{\tilde{E}_i\} = N(E_i) \qquad , \qquad (3.2)$$

This unfolded level sequence $\{\tilde{E}_i\}$ is obviously dimensionless and has a constant average spacing of 1, but the actual spacing exhibit frequently strong fluctuation. The nearest neighbor level spacing is defined as $s_i = (\tilde{E}_{i+1}) - (\tilde{E}_i)$. The distribution $P(s)$ will be in such a way in which $P(s)ds$ is the probability for the $s_i$ to lie within the infinitesimal interval $[s, s+ds]$. For nuclear systems with time reversal symmetry which spectral spacing follows Gaussian Orthogonal Ensemble (GOE) statistics, the NNS probability distribution function is well approximated by Wigner distribution [17]

$$P(s) = \frac{1}{2}\pi s e^{-\frac{\pi s^2}{4}} \qquad , \qquad (3.3)$$

While exhibits the chaotic properties of spectra. On the other hand, the NNSD of systems with regular dynamics is generically represented by Poisson distribution

$$P(s) = e^{-s} \qquad , \qquad (3.4)$$

Different investigations have been accomplished on nuclear system's spectra, proposed intermediate situation of spectral statistics between these limits. To compare the spectral statistics with regular and chaotic limits quantitatively, different distribution functions have been proposed [31-32]. One of popular distribution is Abul-Magd distribution [24] which is derived by assuming that, the energy level spectrum is a product of the superposition of independent subspectra, which are contributed respectively from localized eigenfunctions onto invariant (disjoint) phase space. This distribution is based on the Rosenzweig and Porter random matrix model. The exact form of this model is complicated and its simpler form is proposed by Abul-Magd *et al* in Ref.[24] as:

$$P(s,q) = [1 - q + q(0.7 + 0.3q)\frac{\pi s}{2}] \times \exp(-(1-q)s - q(0.7 + 0.3q)\frac{\pi s^2}{4}) \qquad , \qquad (3.5)$$

Where interpolates between Poisson ($f = 0$) and Wigner ($f = 1$) distributions. In common considerations, one can handle a least square fit (LSF) of Abul-Magd distribution to sequence while the value of distribution's parameter describes the chaotic or regular dynamics [18]. The LSF-based estimated values have some unusual uncertainties and also exhibit more deviation to chaotic dynamics. Consequently, it is almost impossible to carry out any reliable statistical analysis in some sequences. Recently [25], we have employed the Maximum Likelihood Estimation (MLE) technique to estimate every distribution's parameter which provides more precisions with low uncertainties, i.e. estimated values yield accuracies which are closer to Cramer-Rao Lowe Bound (CRLB). Also, this technique yields results which are almost exact in all sequences, even in cases with small sample sizes, where other estimation methods wouldn't achieve the appropriate results. Consequently, we analyzed the spectral statistics of considered systems with about 10 or more samples in each sequence with more precision. The MLE estimation procedure has been described in detail in Ref [25]. Here, we outline the basic ansatz and summarize the results.

- **The ML-based results for Abul-Magd distribution**

The MLE method provides an opportunity for estimating exact result with minimum variation. In order to estimate the parameter of distribution, Likelihood function is considered as product of all $P(s)$ functions,

$$L(q) = \prod_{i=1}^{n} P(s_i) = \prod_{i=1}^{n} [1-q+q(0.7+0.3q)\frac{\pi s_i}{2}] \, e^{-(1-q)s_i - q(0.7+0.3q)\frac{\pi s_i^2}{4}} \quad , \quad (3.6)$$

Then, with taking the derivative of the log of likelihood function (3.6) respect to its parameter ($q$) and set it to zero, i.e., maximizing likelihood function, the following relation for desired estimator (see Appendix (C) of Ref.[25] for more details) is obtained

$$f : \sum \frac{-1+(0.7+0.6q)\frac{\pi s_i}{2}}{[1-q+q(0.7+0.3q)\frac{\pi s_i}{2}]} + \sum s_i + (0.7+0.6q)\frac{\pi s_i^2}{4} \quad , \quad (3.7)$$

We can estimate "$q$" by high accuracy via solving above equation by Newton-Raphson method which is terminated to the following result,

$$q_{new} = q_{old} - \frac{F(q_{old})}{F'(q_{old})} =$$

$$= q_{old} - \frac{\sum \frac{-1+(0.7+0.6q_{old})\frac{\pi s_i}{2}}{[1-q_{old}+q_{old}(0.7+0.3q_{old})\frac{\pi s_i}{2}]} + \sum s_i - (0.7+0.6q_{old})\frac{\pi s_i^2}{4}}{\sum \frac{[0.3\pi s_i][1-q_{old}+q_{old}(0.7+0.3q_{old})\frac{\pi s_i}{2}] - [-1+(0.7+0.6q_{old})\frac{\pi s_i}{2}]^2}{[1-q_{old}+q_{old}(0.7+0.3q_{old})\frac{\pi s_i}{2}]^2} - \sum 0.15\pi s_i^2} \quad , \quad (3.8)$$

Also we have used the difference of both sides of equation (3.8) to obtain the decreasing of uncertainty for estimated values, namely the CRLB for Abul-Magd distribution is defined as [25]

$$Var(\hat{q}) \geq \frac{1}{MF(q)}$$

Where $M$ represents the number of samples and $F(q)$ is used to describe the Fisher information.

## 4. Numerical results

In the present study, we consider the statistical properties of $SU(3) \leftrightarrow \overline{SU(3)}$ transitional region. $\chi$ is considered as control parameters in our description and different values of it, exhibit systems in dynamical symmetry limits, i.e. $\chi = 0, +\sqrt{7}/2, -\sqrt{7}/2$ for $O(6), SU(3)$ and $\overline{SU(3)}$, respectively. On the other hand, the variation of $\chi$ between these limits, describe the transitional region. Consequently, we have considered the $\chi \in [+\sqrt{7}/2, -\sqrt{7}/2]$ region with step length $\chi = 0.1$ to describe $SU(3) \leftrightarrow \overline{SU(3)}$ transitional region. Also, we have examined the same analysis with $\chi = 0.01$ where any significant change didn't appear in results. With employing the eigenvalues of transitional Hamiltonian (2.7) determined by different values of $\chi$, all $0^+$ to $8^+$ levels are determined for systems with total boson number $N = 3$ and 4 in the energy region below $\leq 4$ *Mev* with arbitrary value for $\eta$. All levels are measured from the ground state $0_1^+$ and are normalized to the first excited state $2_1^+$ while as seen from Eq.(2.7), are parameter free. In this approach we achieved 10 levels for systems with $N = 3$ and 18 levels for systems with $N = 4$ while have used to construct sequences by unfolding processes. Then, with employing the MLE technique, the parameter of Abul-Magd distribution estimated by converging value of iteration Eq.(3.8) where as an initial value we have chosen the value of parameter obtained by LSF method.

Since the investigation of the majority of short sequences yields an overestimation about the degree of chaoticity measured by the "$q$" ( Abul-Magd distribution's parameter), therefore, we wouldn't concentrate only on the implicit value of " $q$ " and examine a comparison between the amounts of " $q$ " for different $\chi$ values. It means, the smallest values of "$q$" explain more regular dynamics and vice versa.

The ML-based estimated "q" values are listed in Table5 where the variations of this quantity for considered systems are presented in Figure2. The matrix elements of quadrupole operator in transitional Hamiltonian (2.1) and consequently, the energy spectra of considered systems, i.e. Eq.(2.7), are symmetric with respect to the control parameter" $\chi$ "(where only even powers of $\chi$ appeared in them). The symmetric variations of chaocity degrees for considered systems with respect to $\chi$ are in agreement with this property of spectra.

Table5. The ML-based estimated "q" values (Abul-Magd distribution parameter) for systems with different $\chi$ and $N$ values.

| $\chi$ | $N=3$ $q$ | $N=4$ $q$ |
|---|---|---|
| -1.3 | $0.59 \pm 0.05$ | $0.55 \pm 0.08$ |
| -1.2 | $0.66 \pm 0.06$ | $0.61 \pm 0.05$ |
| -1.1 | $0.64 \pm 0.09$ | $0.63 \pm 0.11$ |
| -1.0 | $0.62 \pm 0.02$ | $0.60 \pm 0.14$ |
| -0.9 | $0.60 \pm 0.10$ | $0.57 \pm 0.09$ |
| -0.8 | $0.62 \pm 0.08$ | $0.59 \pm 0.10$ |
| -0.7 | $0.63 \pm 0.03$ | $0.62 \pm 0.05$ |
| -0.6 | $0.64 \pm 0.07$ | $0.63 \pm 0.07$ |
| -0.5 | $0.62 \pm 0.08$ | $0.64 \pm 0.02$ |
| -0.4 | $0.61 \pm 0.10$ | $0.61 \pm 0.10$ |
| -0.3 | $0.60 \pm 0.04$ | $0.56 \pm 0.09$ |
| -0.2 | $0.61 \pm 0.09$ | $0.60 \pm 0.06$ |
| -0.1 | $0.59 \pm 0.03$ | $0.51 \pm 0.03$ |
| 0 | $0.51 \pm 0.07$ | $0.46 \pm 0.08$ |
| 0.1 | $0.59 \pm 0.03$ | $0.51 \pm 0.03$ |
| 0.2 | $0.61 \pm 0.09$ | $0.60 \pm 0.06$ |
| 0.3 | $0.60 \pm 0.04$ | $0.56 \pm 0.09$ |
| 0.4 | $0.61 \pm 0.10$ | $0.61 \pm 0.10$ |
| 0.5 | $0.62 \pm 0.08$ | $0.64 \pm 0.02$ |
| 0.6 | $0.64 \pm 0.07$ | $0.63 \pm 0.07$ |
| 0.7 | $0.63 \pm 0.03$ | $0.62 \pm 0.05$ |
| 0.8 | $0.62 \pm 0.08$ | $0.59 \pm 0.10$ |
| 0.9 | $0.60 \pm 0.10$ | $0.57 \pm 0.09$ |
| 1.0 | $0.62 \pm 0.02$ | $0.60 \pm 0.14$ |
| 1.1 | $0.64 \pm 0.09$ | $0.63 \pm 0.11$ |
| 1.2 | $0.66 \pm 0.06$ | $0.61 \pm 0.05$ |
| 1.3 | $0.59 \pm 0.05$ | $0.55 \pm 0.08$ |

The "q" values and also Figure 2, suggest a deviation to less regularity for transitional region in compare to both oblate and prolate dynamical symmetry limits. Also, $\chi=0$ which characterizes the $O(6)$ dynamical symmetry limits explores more regularity in compare to $\chi=+\sqrt{7/2}$ and $-\sqrt{7/2}$ which correspond to oblate and prolate limits, respectively. The apparent regularities of the spectrum in the dynamical symmetry limits, namely $O(6), SU(3)$ and $\overline{SU(3)}$ are governed by the approximate conservations of the quantum number describe the collective degrees of freedom. Our results may be interpreted that the coupling between the single particle and collective degrees of freedom is weaker in oblate limit than prolate limit. Also, this result proposes similar spectral statistics as the prediction of Abul-Magd *et al* in Refs.[33] where propose systems provide evidences for $SU(3)$ (or $\overline{SU(3)}$) by $\chi=-\sqrt{7/2}$ (or $\chi=+\sqrt{7/2}$) which have rotational spectra, explore less regular dynamics in compare to $O(6)$ dynamical symmetry limit (by $\chi=0$) which has vibrational spectra, too. It means, nuclei which are spherical (magic or semi magic) ones, are expected to have shell model spectra and consequently, these results confirm the prediction of GOE [34], i.e. the identity of nucleons make impossible to define the rotation for spherical nuclei and the rotation of nuclei contribute to the suppression of their chaotic dynamics.

For systems with $N=3$ and especially 4, the significant variations in the spectral statistics are apparent where for some special values of control parameters, i.e. $\chi\approx\pm0.3$ and $\simeq\pm0.9$, a deviation to more regular dynamics are proposed. One may associate these values of control parameters with the critical values of this transitional region, namely $O(6)\leftrightarrow SU(3)$ (or $\overline{SU(3)}$). Our results may be interpreted that the some special values of control parameter ($\chi$) which describe the level crossing for considered systems, explore a deviation to regular dynamics due to the partial dynamical symmetries in these transitional regions [35]. Also, $\chi=0$ which describe $O(6)$ dynamical symmetry limit and is known as $Z(5)$ critical point of this transitional region [36-37], exhibit more regularity than the predictions for $SU(3)$ (or $\overline{SU(3)}$) dynamical symmetry limit, e.g. we have suggested same results in the Ref. [25] by employing sequences prepared by nuclei which provide empirical evidences for these dynamical symmetry limits.

## 5. Summary and conclusion

In this paper, the spectral statistics of $SU(3)\leftrightarrow\overline{SU(3)}$ transitional region was described in NNSD statistics framework. In the parameter free approach, energy spectra have determined by using the $SO(6)$ representation of eigenstates for systems with total boson number $N=3$ and 4. By employing the MLE technique to estimate with more accuracy, the parameter of Abul-Magd's distribution was estimated where proposed a deviation to less regularity for transitional region between dynamical symmetry limits, namely $O(6)\leftrightarrow SU(3)$ (or $\overline{SU(3)}$). These results may be interpreted a less regular dynamics for transitional region due to symmetry broken or a combination of different symmetries in this regions. Also, some deviations to regularity may caused by partial dynamical symmetries in these regions while $\chi=0$ which describe $O(6)$ dynamical symmetry limits, i.e. $Z(5)$ critical point, explore more regular dynamics. Works in these directions are in progress.

# Figure Caption

**Figure1.** The extend Casten triangle [14], represents different dynamical symmetries of IBM as open circles.

**Figure2(color online).** The variation of chaocity degrees (Abul-Magd distribution's parameter) for considered systems versus control parameter ($\chi$).

Figure1.

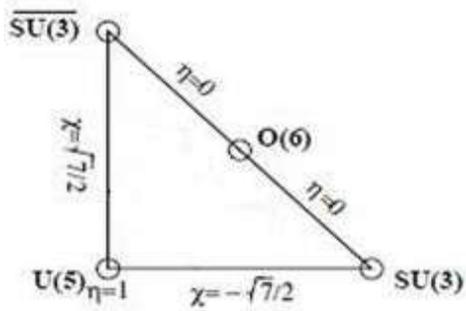

Figure2.

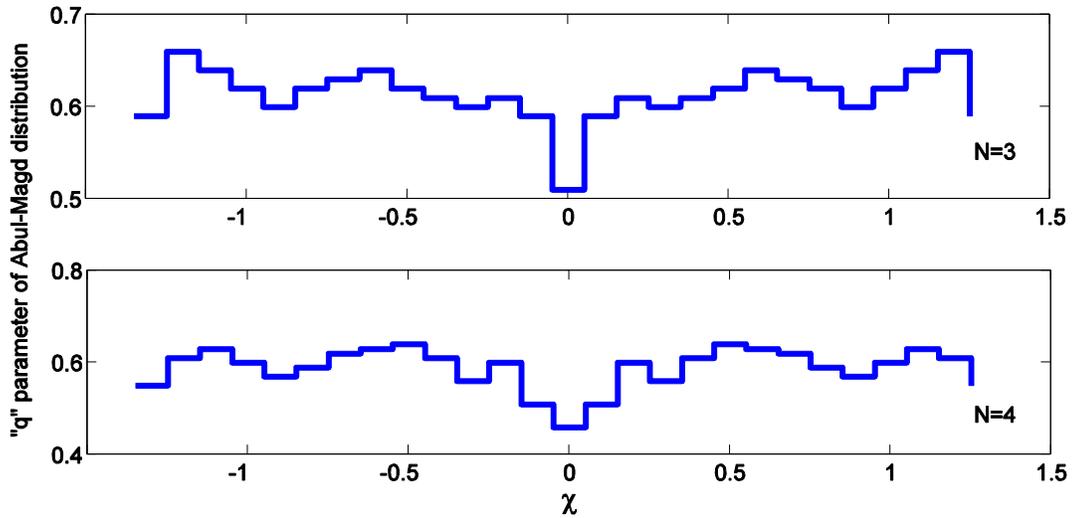